\documentclass{article}

\usepackage{PRIMEarxiv}

\usepackage[utf8]{inputenc} 
\usepackage[T1]{fontenc}    
\usepackage{hyperref}       
\usepackage{url}            
\usepackage{booktabs}       
\usepackage{amsfonts}       
\usepackage{nicefrac}       
\usepackage{microtype}      
\usepackage{lipsum}
\usepackage{fancyhdr}       
\usepackage{graphicx}       
\usepackage{natbib}
\usepackage{amsmath}
\usepackage{siunitx}
\usepackage{xcolor} 
\graphicspath{{media/}}     

\title{AIFS-COMPO : A Global Data-Driven Atmospheric Composition Forecasting System
}

\author{
  Paula Harder, Johannes Flemming, Mihai Alexe, Gert Mertes, Baudouin Raoult, Matthew Chantry \\
  European Centre for Medium-Range Weather Forecasts (ECMWF)\\
}

\begin{document}
\maketitle

\begin{abstract}
We introduce AIFS-COMPO, a skilful medium-range data-driven global forecasting system for aerosols and reactive gases. Building on the ECMWF Artificial Intelligence Forecast System (AIFS), AIFS-COMPO employs a transformer-based encoder–processor–decoder architecture to jointly model meteorological and atmospheric composition variables. The model is trained on Copernicus Atmosphere Monitoring Service (CAMS) reanalysis, analysis, and forecast data to learn the coupled dynamics of weather, emissions, transport, and atmospheric chemistry. We evaluate AIFS-COMPO against a range of atmospheric composition observations and compare its performance with the operational CAMS global forecasting system IFS-COMPO. The results show that AIFS-COMPO achieves comparable or improved forecast skill for several key species while requiring only a fraction of the computational resources. Furthermore, the efficiency of the approach enables forecasts beyond the current operational horizon, demonstrating the potential of AI-based systems for fast and accurate global atmospheric composition prediction.
\end{abstract}

\section{Introduction}
Atmospheric composition (AC) forecasting is essential for understanding and predicting the ground level concentrations of air pollutants such as particulate matter, ozone, and nitrogen dioxide, which have harmful effects on human health linked to respiratory and cardiovascular diseases. By providing early warnings of poor air quality and informing about the evolution of air pollution episodes these forecasts help individuals, healthcare systems, and policymakers take preventive measures to reduce exposure. Operational forecasting systems, such as those provided by the Copernicus Atmosphere Monitoring Service (CAMS)\citep{TheCopernicusAtmosphereMonitoringServiceFromResearchtoOperations}, deliver global and regional analyses and forecasts of atmospheric composition that support air-quality management and policies. Beyond health, atmospheric composition also influences ecosystems and climate processes. 
However, accurately forecasting atmospheric composition remains challenging due to the complexity of the chemical and physical processes involved, the high computational demands of running detailed atmospheric chemistry models, as well as the lack of sufficient observations to constrain the initial conditions motivating the development of more efficient AI-based approaches.

Recently, AI models have achieved impressive results in the field of numerical weather prediction (NWP) \citep{keisler2022forecastingglobalweathergraph,lam2023graphcastlearningskillfulmediumrange,lang2024aifsecmwfsdatadriven}. These models have shown that data-driven approaches can rival or even surpass traditional numerical models while requiring substantially fewer computational resources. The foundation model Aurora \citep{bodnar2025aurora} is one of the first models that, beyond NWP, also demonstrates a data-driven approach to global atmospheric composition forecasting. Atmospheric composition forecasting poses additional challenges for AI compared to weather prediction. In contrast to the relatively well-observed meteorological state variables, atmospheric composition is governed by complex chemical processes driven by highly heterogeneous (in time and space) emissions from natural and anthropogenic sources. The lack and heterogeneity of observation in some parts of the world posing an additional issue. Chemical processes lead to strong spatial variability and non-linear interactions across multiple temporal scales, making global atmospheric compound distributions more difficult to learn. 

A growing body of research is now emerging to apply machine learning (ML) across different components and spatial scales of atmospheric composition modelling. At the local scale, ML approaches are widely used for statistical downscaling \citep{gmd-15-6677-2022, SHETTY2025120363} and high-resolution mapping of pollutants by combining observations, satellite data, and chemical transport model outputs \citep{egusphere-2026-1109}. Regional AI-based chemistry transport models such as Zeeman \citep{pang2025zeemandeeplearningregional} and BiXiao \citep{egusphere-2025-5589}, aim to learn the coupled dynamics of emissions, chemistry, and transport directly from data. Another line of work focuses on emulating individual model components or ensembles in order to accelerate computationally expensive processes within existing modelling frameworks; for example, EnsAI \citep{sitwell2026ensaiemulatoratmosphericchemical} generates atmospheric chemical ensembles orders of magnitude faster than traditional simulations. Finally, recent studies, alongside Aurora, explore global fully data-driven models for atmospheric composition, such as AI-GAMFS, that predicts aerosol properties worldwide \citep{gui2026aerosol}.

In this work, we introduce AIFS-COMPO, an extension of ECMWF's AIFS system to atmospheric composition, and the first AI model to provide a global three-hourly forecast of a wide range of key atmospheric composition variables. AIFS-COMPO builds on the transformer-based encoder–processor–decoder architecture of the original AIFS model \citep{lang2024aifsecmwfsdatadriven} and is implemented within the  Anemoi\footnote{\url{https://github.com/ecmwf/anemoi}} framework for data-driven weather and climate modelling. The model jointly predicts meteorological and atmospheric composition variables on a global 80\,km grid with a temporal resolution of three hours, enabling the representation of diurnal variability and pollution peaks at specific times of day. In total, it forecasts 187 variables, partially as surface or integrated quantities and partially across multiple pressure levels. AIFS-COMPO is trained on CAMS reanalysis EAC4\footnote{Available at \url{https://ads.atmosphere.copernicus.eu}}, as well as operational analysis and forecast data. 

We evaluate AIFS-COMPO against a wide range of independent atmospheric composition observations and compare its performance with the operational CAMS forecasting system IFS-COMPO. This evaluation against independent observations is essential to assess real-world forecast skill beyond consistency with training data. Our results show that AIFS-COMPO achieves similar or improved skill for several key atmospheric composition variables. Once trained, AIFS-COMPO requires only a fraction of the computational resources of the numerical model, producing a 3-hourly, 5-day global forecast in 50\,s on a single GPU, compared to roughly 1000\,s on 8000 CPUs for IFS-COMPO. Furthermore, the efficiency of the data-driven approach enables forecasts beyond the current 5-day operational horizon. We demonstrate the quality of these extended forecasts through case studies of large-scale atmospheric composition phenomena, including the prediction of the development of the Antarctic ozone hole.

\section{Data}

The data used in this study consist of atmospheric composition modelling data used to train AIFS-COMPO and atmospheric composition observations used for verification. The modelling data include reanalysis, analysis, and forecast products from the CAMS which contain both standard NWP variables and atmospheric composition (AC) variables.

\subsection{Variables}

In addition to standard meteorological variables from numerical weather prediction, AIFS-COMPO includes a range of atmospheric composition variables describing aerosols and reactive gases. Both NWP and AC variables are treated as prognostic quantities and are therefore used as inputs and predicted outputs of the model. Variables are defined either on pressure levels or as surface or total-column quantities. In addition, several temporal and static features are included, such as the day of year, local time, and static geographic information including orography. The full list of variables is shown in Table~\ref{tab:vars}.

\begin{table}[htb]
 \caption{Variables used in AIFS-COMPO. Pressure levels used here are 50,
100, 150, 200, 250,
300, 400, 500, 600,
700, 850, 925, 1000.}
  \centering
  \begin{tabular}{l|l|l|l}
    \toprule
    Type & Field    & Leveltype     & Input/Output \\
    \midrule
    NWP variables &Geopential, horizontal and vertical wind components & Pressure level  & Both   \\
     & specific humidity, wind speed, temperature &   &   \\
     \midrule
    NWP variables  & Surface pressure, mean sea-level, 2m temperature  & Surface &  Both    \\
 & skin temperature, 2m dewpoint temperature &  &    \\
  & 10 m horizontal and vertical wind components &  &      \\
   & total column water and water vapour &  &      \\
   \midrule
   AC variables & Mass mixing ratios: ozone, sulphure dioxide, nitrogen dioxide & Pressure level  & Both   \\
         & nitrogen oxide, carbon monoxide &       & Both  \\
          \midrule
        AC variables & Aerosol optical depths at 550nm: Total, dust, sea salt & Surface  & Both   \\
         & sulphate, organic matter, black carbon &       &  \\
          & Particulate matter at 1\textmu m, 2.5\textmu m, 10\textmu m &       &  \\
           & Total column: ozone, sulphure dioxide, nitrogen dioxide  &       &  \\
           & carbon monoxide &       &   \\
           & Surface: ozone, sulphure dioxide, nitrogen dioxide  &       &  \\
           & nitrogen oxide, carbon monoxide &       &   \\
          \midrule
      Add. features & Land-sea mask, orography, std of sub-gridscale orography & Surface  & Input   \\
      & slope of sub-gridscale orography, insolation,  &&  \\
       & cos/sin of latitude, longitude, time of day, day of year & & \\
        & cos solar zenith angle & & \\
    \bottomrule
  \end{tabular}
  \label{tab:vars}
\end{table}

The most important variables for air quality are pollutants near the surface, including particulate matter (PM), ozone (O$_3$), nitrogen dioxide (NO$_2$), nitrogen oxide (NO), sulphur dioxide (SO$_2$), and carbon monoxide (CO). Particulate matter is classified into three size ranges according to particle diameter: PM$_{10}$, PM$_{2.5}$, and PM$_1$, corresponding to particles with diameters smaller than 10, 2.5, and 1\textmu m, respectively. Aerosol optical depth (AOD) measures the extinction of sunlight by aerosol particles in the atmosphere; here we consider AOD at the commonly used wavelength of 550nm.
Aerosols originate from a variety of sources and consist of different chemical compounds, including dust, sea salt, sulphate, black carbon, and organic matter. These aerosol types have distinct physical and radiative properties and therefore different impacts on atmospheric processes. For this reason, they are represented separately in the model.

\subsection{Training Data} 

For training and validation we use several CAMS model datasets, described below.

\subsubsection{Reanalysis Data}

The primary dataset used for training is the fourth-generation CAMS atmospheric composition reanalysis (EAC4) \citep{acp-19-3515-2019}, which also includes standard meteorological variables used in numerical weather prediction. The term reanalysis refers to the application of a consistent data assimilation system over a historical period, combining model simulations with observations to provide a best estimate of the past state of the atmosphere. EAC4 was produced by assimilating satellite retrievals of aerosols, ozone, carbon monoxide and nitrogen dioxide as well as of stratospheric ozone profiles. The data lies on a reduced Gaussian grid, which is a grid designed to maintain approximately uniform horizontal spacing between grid points while reducing the number of grid points towards the poles. The version used in this study is the N128 grid, which contains 128 latitude lines between the equator and the pole and corresponds to a horizontal resolution of approximately 80~km.

\subsubsection{Operational Data} \label{oper_data}

To incorporate recent model updates and extend the training data volume, we additionally include operational CAMS analysis and forecast data covering the period from 2019 to 2023. While reanalysis corresponds to a retrospective application of a fixed data assimilation system over a historical period, the analysis represents the same concept in an operational, continuously updated setting using the latest model version. The operational dataset used here consists of a combination of analysis and forecast fields, as the CAMS atmospheric composition analysis is produced only every 6~hours. To obtain a consistent 3-hourly dataset, the missing intermediate times are filled using short-range forecasts. Specifically, 3-hour forecasts are used for the times 03:00 and 15:00~UTC, and 9-hour forecasts are used for 09:00 and 21:00~UTC, since the operational forecasts are initialized every 12~hours. The operational data are originally produced on a higher-resolution grid with approximately 40~km spacing. To ensure consistency with the reanalysis data, the operational fields are interpolated to the N128 grid used for training.

The AIFS-COMPO model was trained with data until the end of 2023 with 2024 used as the test year.

\subsection{Observations} \label{obs_data}

As a key point of this work, independent observations are used for verification of the forecasts but not used during training. The observational datasets consist of ground-based measurements from several monitoring networks that provide information on key atmospheric composition variables. For more detailed information on each dataset we refer to \citet{eskes2025cams_eqc_observations}.

\begin{itemize}

\item \textbf{AERONET (Aerosol Robotic Network)}: A global network of ground-based sun photometers measuring spectral aerosol optical depth (AOD) \citep{HOLBEN19981}. The network consists of about 400 stations, mainly located over land, with particularly dense coverage in North America and Europe.

\item \textbf{AirNow}: Surface air quality observations for North America are obtained from the AirNow partnership and Environment Canada. The dataset includes routine measurements from hundreds of stations (approximately 900 for O$_3$, 200 for NO$_2$, 220 for PM$_{10}$, and 670 for PM$_{2.5}$).

\item \textbf{AQ e-reporting}: European surface air quality observations collected by the European Environment Agency (EEA) and reported by the European countries with respect to the Ambient Air quality Directive provisions\footnote{Formerly known as Air Quality Database (AirBase)}. For validation, rural background stations classified following \citet{JOLY2012111} are used, resulting in approximately 665 stations across Europe.

\item \textbf{China AQ}: A national air quality monitoring network operated by the China National Environmental Monitoring Center (CNEMC), comprising more than 1,700 stations across major Chinese cities measuring key pollutants.

\item \textbf{Ozonesondes}: Vertical ozone profile measurements obtained from balloon-borne ozonesondes. The data are compiled from several networks and data centres, including WOUDC, SHADOZ, NDACC, and the MATCH campaigns.

\end{itemize}

\section{Model}

Following the success of large-scale machine learning architectures for numerical weather prediction (NWP), such as vision transformers \citep{bi2022panguweather3dhighresolutionmodel, bodnar2025aurora} and graph neural networks \citep{keisler2022forecastingglobalweathergraph, lam2023graphcastlearningskillfulmediumrange}, AIFS-COMPO  builds on the AIFS model architecture \citep{lang2024aifsecmwfsdatadriven}. While the overall architecture remains largely unchanged, several modifications are introduced, primarily in the treatment of additional atmospheric composition variables and in the training procedure.

\subsection{Architecture}

AIFS-COMPO implements an encoder–processor–decoder model architecture. The encoder and decoder are based on graph neural networks (GNNs) with multi-head graph attention. The encoder maps the input data from the N128 (re)analysis grid (approximately 80km horizontal resolution) to a O48 octahedral latent grid (approximately 210km horizontal resolution), on which the processor operates, and the decoder projects the processed features back to the original grid. The processor consists of a 16-layer pre-norm transformer with shifted window attention, calculated across latitude bands and GELU activations \citep{lang2024aifsecmwfsdatadriven}.

To accommodate the extended set of atmospheric composition variables, the number of input and output channels is increased to 1600 compared to the original AIFS model which includes 1024 channels. For atmospheric composition variables, a ReLU-based output constraint is applied to enforce non-negativity of the predictions, following \citet{moldovan2025updateecmwfsmachinelearnedweather}.


\subsection{Training}

AIFS-COMPO is trained to predict 3-hourly time steps in an auto-regressive manner. This temporal resolution is chosen to match the operational CAMS forecast frequency and, crucially, to capture diurnal variability and short-term fluctuations in atmospheric composition, such as pollution peaks occurring at specific times of day. The model takes as input two consecutive states, at times $t_0$ and $t_{-3\mathrm{h}}$, and predicts the state at $t_{+3\mathrm{h}}$. Longer forecasts are obtained by iteratively feeding model predictions back as input, a procedure referred to as rollout.

Training is performed in three successive stages, summarized in Table~\ref{tab:training-phases}. In particular, finetuning on operational data plays a critical role, as it exposes the model to the most recent system configuration and to differences between reanalysis and the current operational analysis and forecast data. This step is therefore essential to bridge the gap between retrospective training data and real-time forecasting conditions, and cannot be replaced by reanalysis pretraining alone. To further improve multi-step forecast skill and stability, rollout is incorporated during the final stage of training. A comparison of performance against observations for the different training stages is provided in the appendix (Figure \ref{obs-stages}), highlighting the substantial gains obtained from the additional training phases beyond reanalysis pretraining. The model is optimized using the AdamW optimizer\footnote{with $\beta_1 = 0.9$ and $\beta_2 = 0.95$}, combined with a cosine learning rate scheduler and a linear warm-up over the first 1000 steps of each stage; the learning rate values for each stage are specified in Table~\ref{tab:training-phases}.

Input and output variables are normalized to zero mean and unit variance. Atmospheric composition variables subject to ReLU constraints are scaled by their standard deviation only, preserving non-negativity. Static input features, such as orography, are normalized using min–max scaling.

The loss function is an area-weighted mean squared error (MSE), with additional weights depending on the variables. Initial loss weighting for meteorological variables are inherited from AIFS, while atmospheric composition variables are initially assigned equal weights and were subsequently tuned during the model design exploration phase based on validation performance. 
The configuration used here assigns relatively lower weights to aerosol optical depth and particulate matter, and higher weights to ozone. These weights are kept constant across all training stages.

\begin{table}[htb]
 \caption{Training stages for AIFS-COMPO.}
  \centering
  \begin{tabular}{l|l|l|l|l|l}
    \toprule
    Stage & Data & Years & Iterations & Learning rate range & Rollout \\
    \midrule
    1. Pretraining & EAC4 & 2003--2023 & 250k & $5 \cdot 10^{-4}$ -- $3 \cdot 10^{-7}$ & 1 \\
    2. Operational finetuning & Operational & 2019--2023 & 60k & $3 \cdot 10^{-4}$ -- $3 \cdot 10^{-7}$ & 1 \\
    3. Rollout finetuning & Operational & 2019--2023 & $1.6k  \cdot 24$ & $1.46 \cdot 10^{-4}$ -- $3 \cdot 10^{-7}$ & 1--24 \\
    \bottomrule
  \end{tabular}
  \label{tab:training-phases}
\end{table}

\paragraph{Reanalysis Pretraining}

In the first stage, the model is trained on the EAC4 reanalysis dataset covering the period 2003–2023 for 250\,000 iterations. This stage leverages the large temporal coverage and diversity of the reanalysis data, providing access to two decades of heterogeneous atmospheric conditions. As such, it enables the model to learn robust, large-scale relationships between meteorology and atmospheric composition across a wide range of regimes. The relatively high initial learning rate ($5 \cdot 10^{-4}$) and large number of iterations reflect the central role of this stage in establishing a strong general representation.

\paragraph{Operational Finetuning}

In the second stage, the model is fine-tuned using operational CAMS analysis and forecast data (see Section~\ref{oper_data}) for the period 2019–2023. This step is essential to adapt the model to the most recent system configuration, as there are substantial differences between the reanalysis dataset and the current operational analysis and forecast data. Finetuning therefore bridges this gap and ensures consistency with real-time forecasting conditions. Compared to pretraining, a lower initial learning rate ($3 \cdot 10^{-4}$) and fewer iterations (60k) are used, reflecting both the smaller dataset size and the more targeted nature of this adaptation phase.

\begin{figure*}[htb]
\begin{center}
\includegraphics[width=16cm]{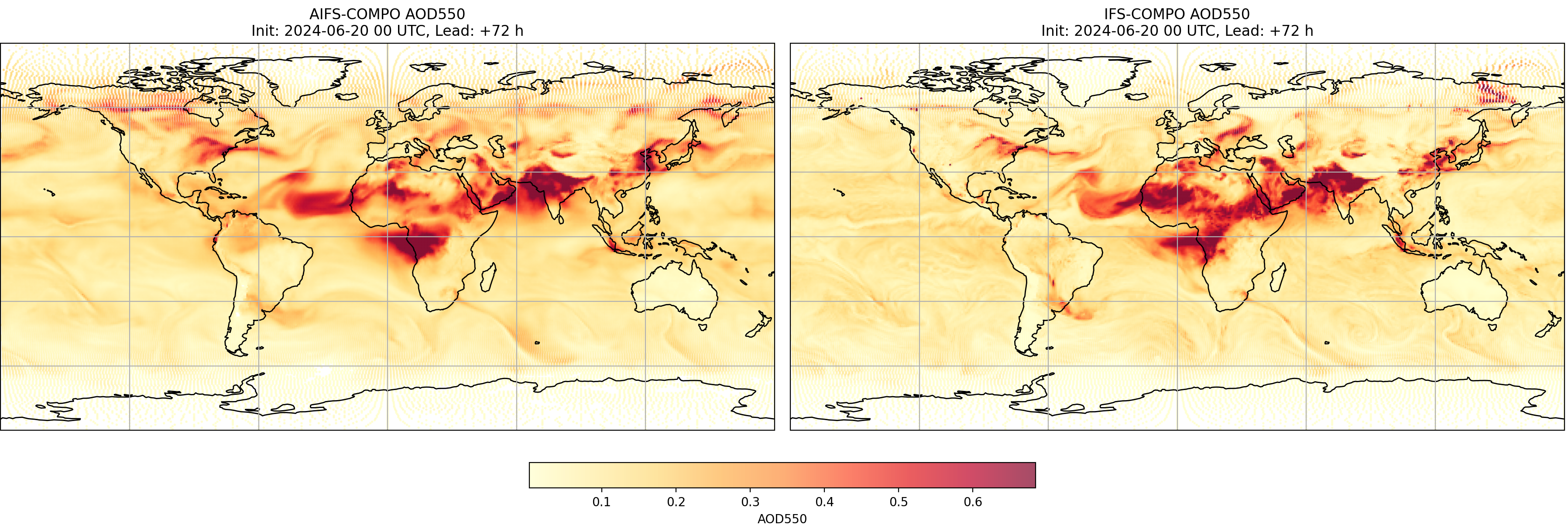}
\caption{A random sample of the day 3 forecast of AIFS-COMPO and IFS-COMPO for total AOD at 550nm.}
\label{obs-aod-aifs-compo-ifs-compo}
\end{center}
\end{figure*}

\paragraph{Rollout Finetuning}

In the final stage, rollout finetuning is applied to improve stability and accuracy for multi-step forecasts. Training is again performed on the operational dataset (2019–2023), while gradually increasing the rollout length from 1 up to 24 steps, keeping each rollout length for one epoch, i.e. about 1,600 iterations. This corresponds to extending the effective forecast horizon during training to up to 3 days. A reduced initial learning rate ($1.46 \cdot 10^{-4}$) is used to ensure stable optimization during long rollouts. We observe that this stage leads to further improvements, particularly at longer lead times, by reducing error accumulation and enhancing temporal consistency.

Training is performed on multiple GPU nodes using a combination of data parallelism and model parallelism (for rollout finetuning). The total training time is about 5 days on 16 GPUs. For pretraining we use an effective batch size of 32 for finetuning a batch size of 16.


\begin{figure*}[htb]
\begin{center}
\includegraphics[width=16cm]{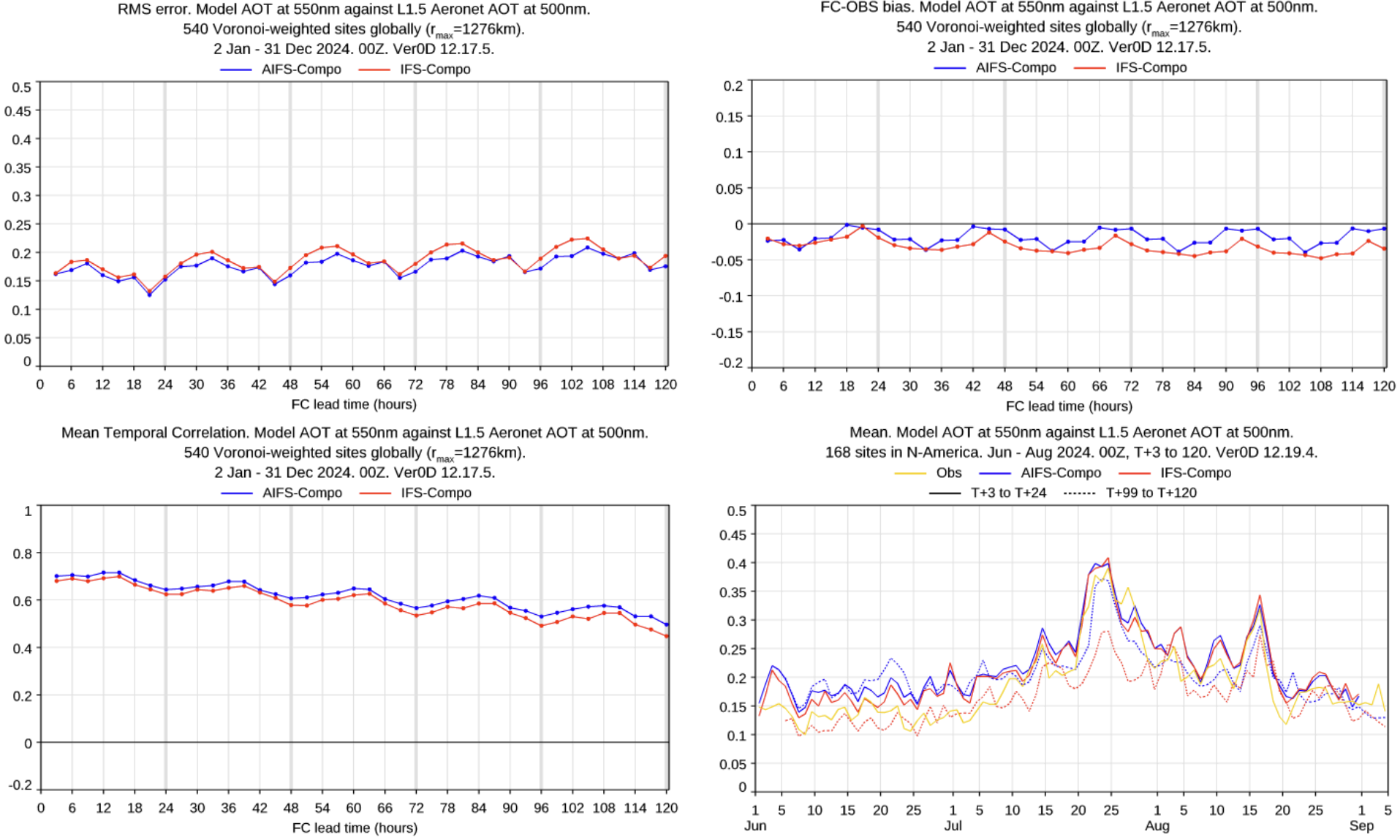}
\caption{AOD prediction performance of IFS-COMPO (red) and AIFS-COMPO (blue) compared against Aeronet observations (yellow). Top left: global RMSE, top right: global bias, bottom left: global temporal correlation, bottom right: a JJA AOD timeseries for North America.}
\label{aod-map}
\end{center}
\end{figure*}

\section{Results}

We compare the skill of AIFS-COMPO against the operational CAMS forecast system, IFS-COMPO\footnote{The IFS-COMPO data used here for comparison against observations are at a higher horizontal resolution of approximately 40\,km.}. The key strength of this evaluation is the extensive use of independent observational datasets, which provide a robust and physically grounded assessment of forecast quality beyond model-internal consistency. Verification is performed against multiple observation networks for aerosols, particulate matter, reactive gases, and ozone (see Section~\ref{obs_data}) over the full year 2024. In addition, we analyse two case studies: the North American wildfire season (JJA 2024) and the Antarctic ozone hole. Since observational verification is only available for a subset of variables, we further include an evaluation against CAMS analysis (Section~\ref{res-ana} and Appendix). Figure~\ref{aod-map} illustrates a representative example of a 3-day AOD forecast, highlighting that AIFS-COMPO captures the large-scale spatial patterns of aerosol distributions comparably to IFS-COMPO. Small-scale features appear slightly smoother, which can be attributed to the lower-resolution training data as well as the tendency of RMSE-optimised models to produce spatially smoothed predictions.

\subsection{Aerosol Optical Depth}

We first evaluate aerosol optical depth (AOD) at 550\,nm using AERONET observations. Figure~\ref{obs-aod-aifs-compo-ifs-compo} shows that AIFS-COMPO consistently achieves lower RMSE, reduced bias, and higher temporal correlation compared to IFS-COMPO. This improvement is also evident in regional time series. For example, during summer 2024 in North America, both AIFS-COMPO and IFS-COMPO accurately capture a strong AOD peak at short lead times. Notably, the 5-day forecast of AIFS-COMPO is still able to reproduce this, whereas the corresponding IFS-COMPO  forecast shows a clear degradation.

\subsection{Particulate Matter}

\begin{figure*}[htb]
\begin{center}
\includegraphics[width=\columnwidth]{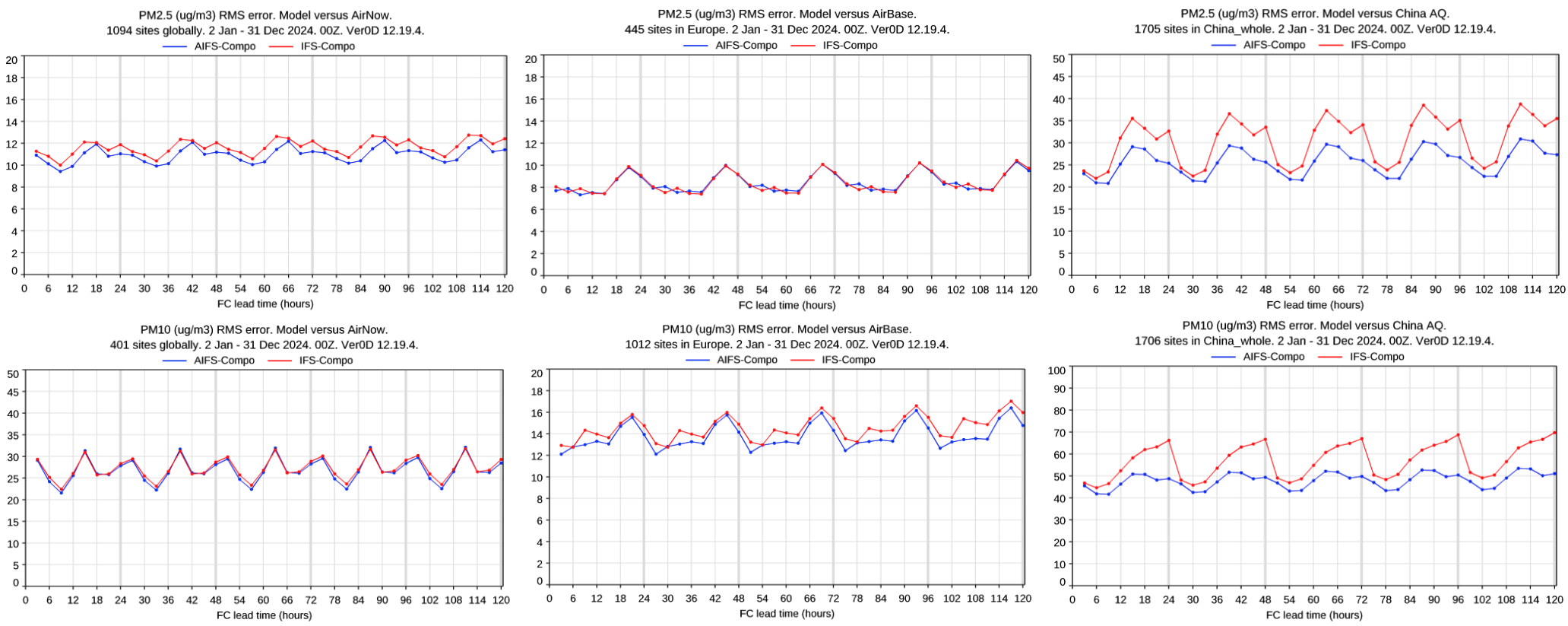}
\caption{Comparison of PM predictions for North America, Europe, and China. First row is showing PM$_{2.5}$, second row is showing PM$_{10}$.}
\label{obs-pms-aifs-compo-ifs-compo}
\end{center}
\end{figure*}

As shown in Figure~\ref{obs-pms-aifs-compo-ifs-compo}, we evaluate particulate matter concentrations for PM$_{2.5}$ and PM$_{10}$ using observations from AirNow (North America), AQ e-reporting database (Europe), and China AQ. AIFS-COMPO achieves lower RMSE than IFS-COMPO in four out of six cases, with the largest improvements observed over China. For PM$_{2.5}$ in Europe and PM$_{10}$ in North America, both models exhibit comparable performance, indicating that the data-driven approach is able to match the skill of the physical model in these regions.

\subsection{Reactive Gases}

\begin{figure*}[htb]
\begin{center}
\includegraphics[width=\columnwidth]{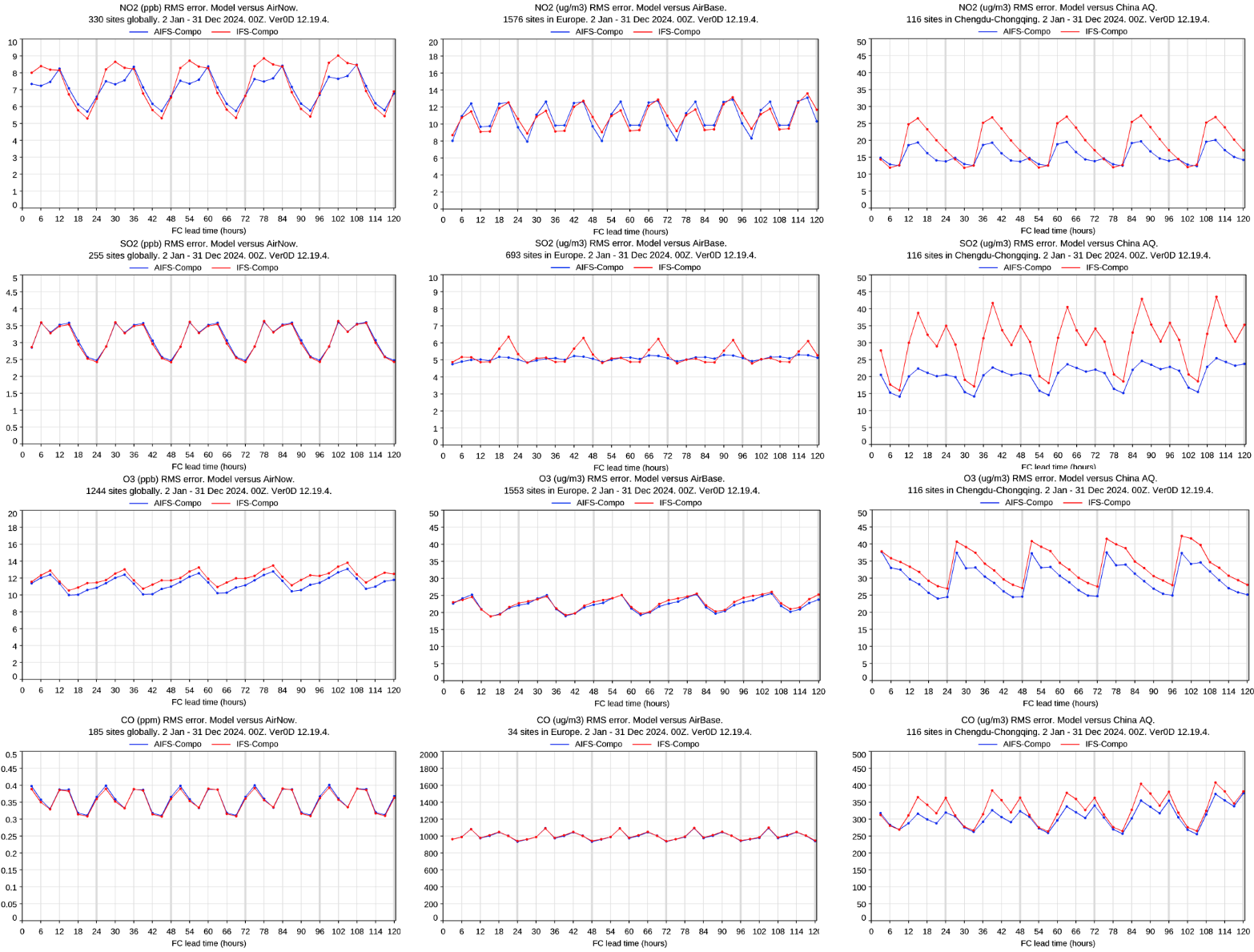}
\caption{Comparison of NO$_2$, SO$_2$, ozone, and CO predictions of AIFS-COMPO (blue) and IFS-COMPO (red) against observations in North America (first column), Europe (second column), and China (third column).}
\label{obs-gases-aifs-compo-ifs-compo}
\end{center}
\end{figure*}

We assess model performance for surface concentrations of reactive gases, including NO$_2$, SO$_2$, CO, and ozone, see Figure \ref{obs-gases-aifs-compo-ifs-compo}. For NO$_2$, SO$_2$, CO, and ozone, evaluation is performed over North America, Europe, and China,. 
Figure~\ref{obs-gases-aifs-compo-ifs-compo} shows that performance varies depending on both species and region, IFS-COMPO and AIFS-COMPO perform similarly or AIFS-COMPO improves over the operational system, with the largest improvememnts over China. For NO$_2$, AIFS-COMPO generally performs similar for North America and Europe, but clearly outperforms IFS-COMPO for China. For SO$_2$, results are more region-dependent: performance is nearly identical over North America, mixed depending on time of the day over Europe, and significantly improved over China. For surface ozone, AIFS-COMPO improves upon IFS-COMPO over North America and China, while achieving comparable performance over Europe.

\subsection{Ozone Profiles}

\begin{figure*}[htb]
\begin{center}
\includegraphics[width=\columnwidth]{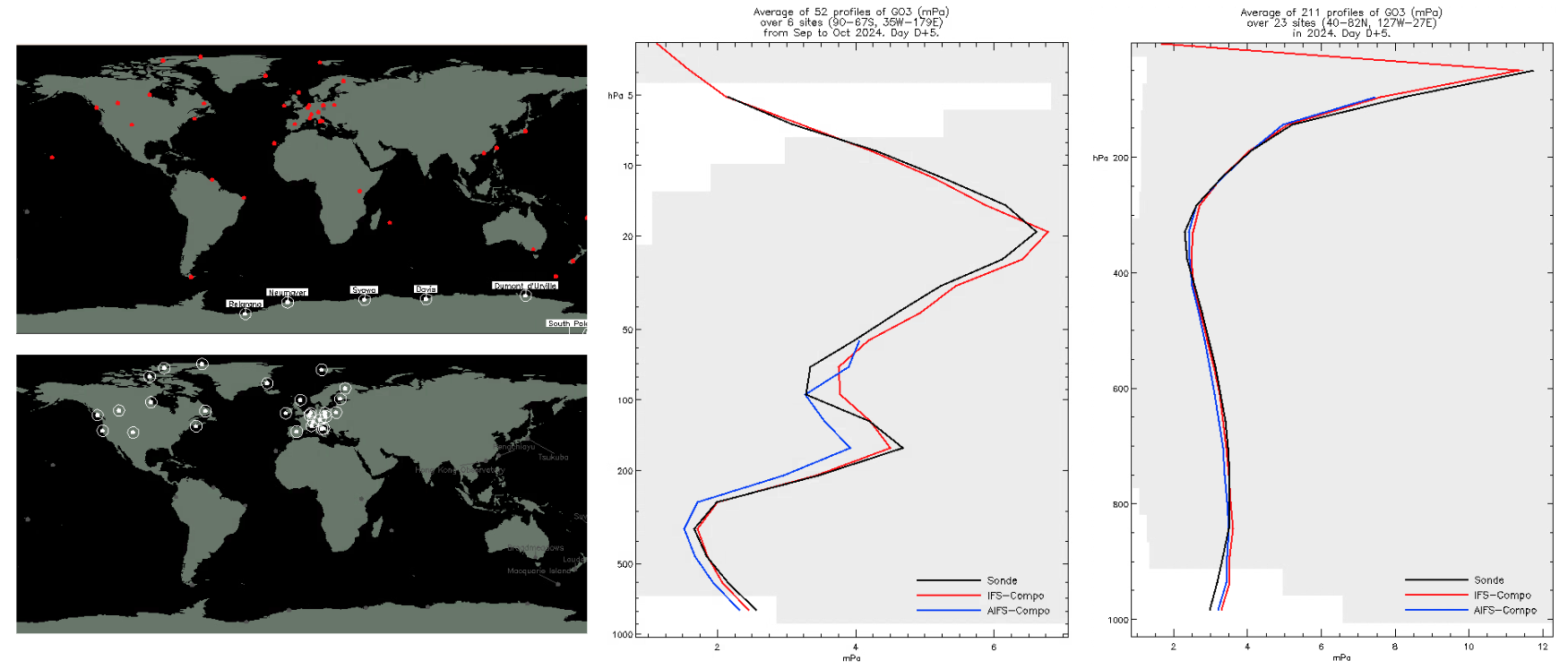}
\caption{Evaluation of ozone profiles. Left: locations of Antarctic stations (top) and North American/European stations (bottom) in white circles. Middle and right: comparison of model predictions, AIFS-COMPO (blue) and IFS-COMP (red) with sonde observations (black). Middle: Using Antarctic stations to capture the ozone hole in September and October, plotted with logarithmic pressure scale. Right: Using all stations of North American/European over a whole year and a linear scale.}
\label{obs-ozone-profiles}
\end{center}
\end{figure*}

Vertical ozone profiles are evaluated using ozonesonde observations. We compare model output on pressure levels up to 50\,hPa against measurements. Two evaluation setups are considered: (i) a full-year analysis over 23 stations in North America and Europe, and (ii) a focused analysis over Antarctic stations to assess ozone hole representation. As visible in Figure \ref{obs-ozone-profiles}, across most pressure levels, AIFS-COMPO captures the vertical structure of ozone well and shows competitive agreement with observations relative to IFS-COMPO.

\subsection{Case Study: Ozone Hole Prediction}

\begin{figure*}[htb]
\begin{center}
\includegraphics[width=10cm]{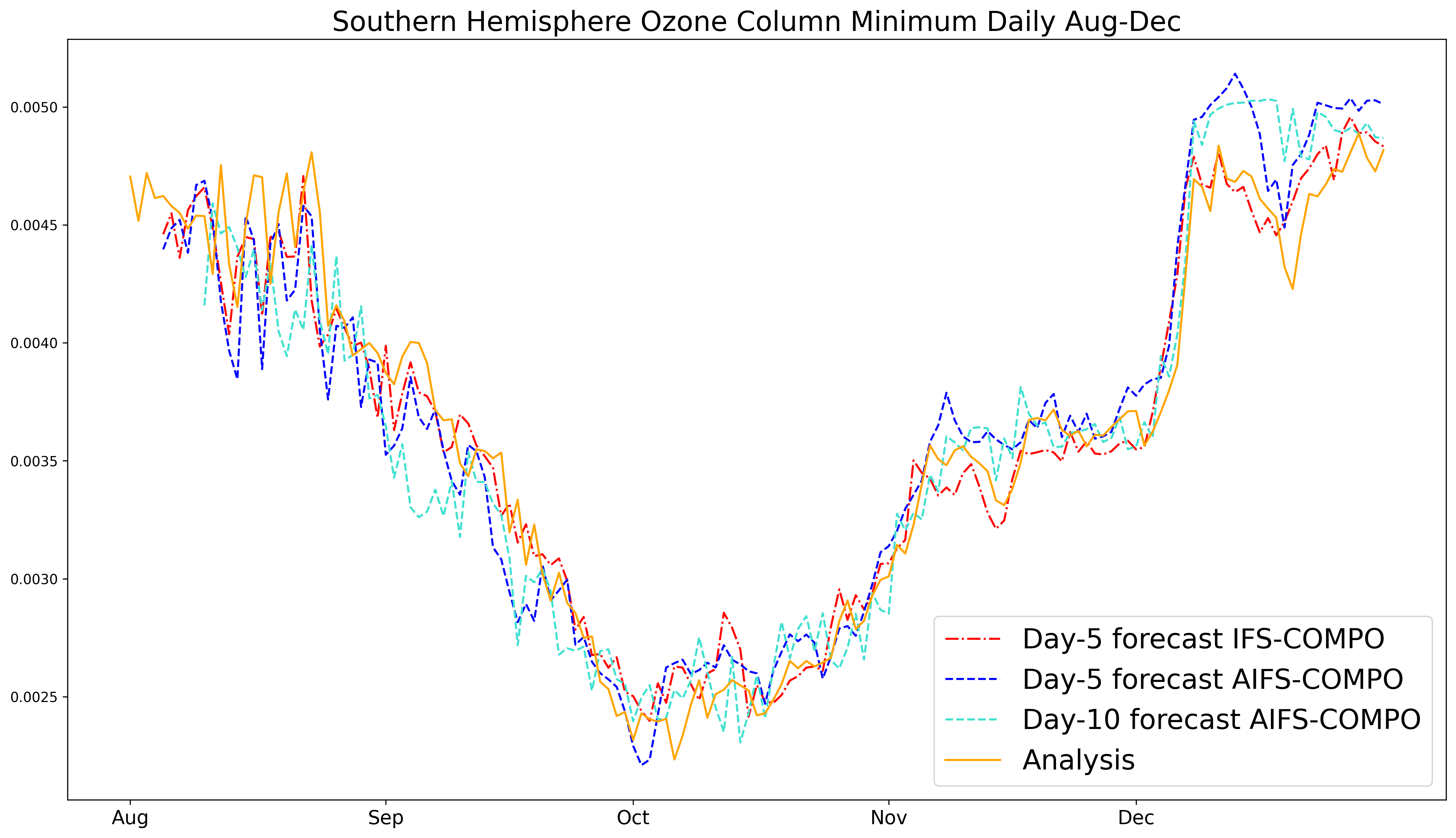}
\caption{Ozone hole over the Southern Hemisphere (August–December 2024), showing analysis, 5-day forecasts from IFS-COMPO and AIFS-COMPO, and a 10-day forecast from AIFS-COMPO.}
\label{obs-ozone-hole}
\end{center}
\end{figure*}

We investigate the ability of AIFS-COMPO to predict the Antarctic ozone hole during the period August–December 2024. Due to its low computational cost, AIFS-COMPO enables extended forecasts beyond the typical 5-day horizon of IFS-COMPO. As shown in Figure~\ref{obs-ozone-hole}, the 5-day forecast of AIFS-COMPO is comparable in quality to the 5-day IFS-COMPO forecast. Importantly, even at 10-day lead time, AIFS-COMPO is able to reliably capture the development and extent of the ozone hole, demonstrating the potential of AI-based systems for longer-range atmospheric composition forecasting.

\subsection{Comparison Against Analysis} \label{res-ana}

We further evaluate model performance against global CAMS analysis data over a full year (see Appendix for details). Overall, AIFS-COMPO exhibits lower RMSE than IFS-COMPO at longer lead times for many variables, although performance varies depending on the species. For aerosol optical depth, AIFS-COMPO outperforms IFS-COMPO after the first forecast day for total AOD and most species. For particulate matter, AIFS-COMPO surpasses IFS-COMPO after 1–2 days for PM$_1$ and PM$_{2.5}$, while PM$_{10}$ remains more challenging. For total column variables, AIFS-COMPO performs slightly worse than IFS-COMPO, particularly for ozone. For pressure-level variables, AIFS-COMPO shows strong improvement compared to IFS-COMPO for many pressure levels and leadtimes, but performance varies across variables, e.g. for stratospheric ozone or near surface NO2 AIFS-COMPO shows still a much larger error than IFS-COMPO with respect to analysis.

\section{Discussion and Conclusion}

This work demonstrates that large-scale machine learning architectures developed for numerical weather prediction, such as AIFS, can be successfully extended to atmospheric composition forecasting, provided that they are trained on robust and reliable atmospheric composition reanalysis datasets that adequately capture the underlying chemical complexity.
With AIFS-COMPO, we introduce the first global, data-driven atmospheric composition model that provides 3-hourly forecasts across a wide range of aerosols and reactive gases, and we show that it achieves performance comparable to—or exceeding—that of the operational CAMS system IFS-COMPO when evaluated against independent observations.

Across multiple observational datasets, AIFS-COMPO matches or outperforms the physical model for key variables such as aerosol optical depth and particulate matter, while showing competitive skill for reactive gases and ozone. At the same time, the model requires only a fraction of the computational resources, enabling substantially faster inference. This efficiency opens up new possibilities for atmospheric composition forecasting, in particular the extension to longer lead times. Notably, we demonstrate that AIFS-COMPO is able to produce skilful forecasts of the Antarctic ozone hole up to 10 days in advance, maintaining comparable quality to the 5-day forecasts of IFS-COMPO. This highlights the potential of AI-based systems to provide earlier warnings for large-scale atmospheric composition events. While the results are encouraging, several limitations remain. In comparisons against CAMS analysis, AIFS-COMPO shows reduced skill at short lead times and for certain variables, particularly ozone and total column quantities. 

A promising direction for future work is the inclusion of emissions as explicit model inputs, as they are a primary driver of atmospheric composition variability. This would also enable scenario-based forecasting at very low computational cost. Another important avenue is the integration of observations into the training process. Given the relatively large initial condition errors in atmospheric composition, learning directly from observations \citep{alexe2024graphdopskilfuldatadrivenmediumrange}, or jointly learning from model data and observational constraints \citep{vaughan2024aardvarkweatherendtoenddatadriven}, could significantly improve forecast accuracy, particularly at short lead times.

Further improvements could be achieved through advances in model design and training, such as alternative normalization strategies \citep{bodnar2025aurora}, improved representation of boundary conditions including surface fluxes and emissions, extension to additional variables (e.g. vertically resolved aerosol properties, greenhouse gases, and radiation fields), and increased vertical and horizontal resolution to match those of current operational systems. Increasing the spatial resolution to match that of IFS-COMPO is also a future target, with results from NWP demonstrating that resolution transfer can be achieved during training \citep{nipen2025regional}. In addition, probabilistic formulations of AIFS-COMPO, for example through training with distribution-based loss functions such as the continuous ranked probability score (CRPS) \citep{lang2024aifscrpsensembleforecastingusing}, could help mitigate the spatial smoothing inherent to RMSE-based optimisation by better representing forecast uncertainty and extremes. The reduced computational cost of AIFS-COMPO also makes it feasible to develop ensemble-based forecasting systems, which are currently standard in NWP but remain computationally prohibitive for atmospheric composition.

\section*{Acknowledgments}
The authors gratefully acknowledge the support of Laurence Rouil, Richard Engelen, and their colleagues at ECMWF for their valuable contributions to this work.

\newpage
\bibliographystyle{plainnat}  
\bibliography{references}  

\newpage
\section*{Appendix}

\subsection*{Evaluation of training stages} \label{stages}

\begin{figure*}[htb]
\begin{center}
\includegraphics[width=\columnwidth]{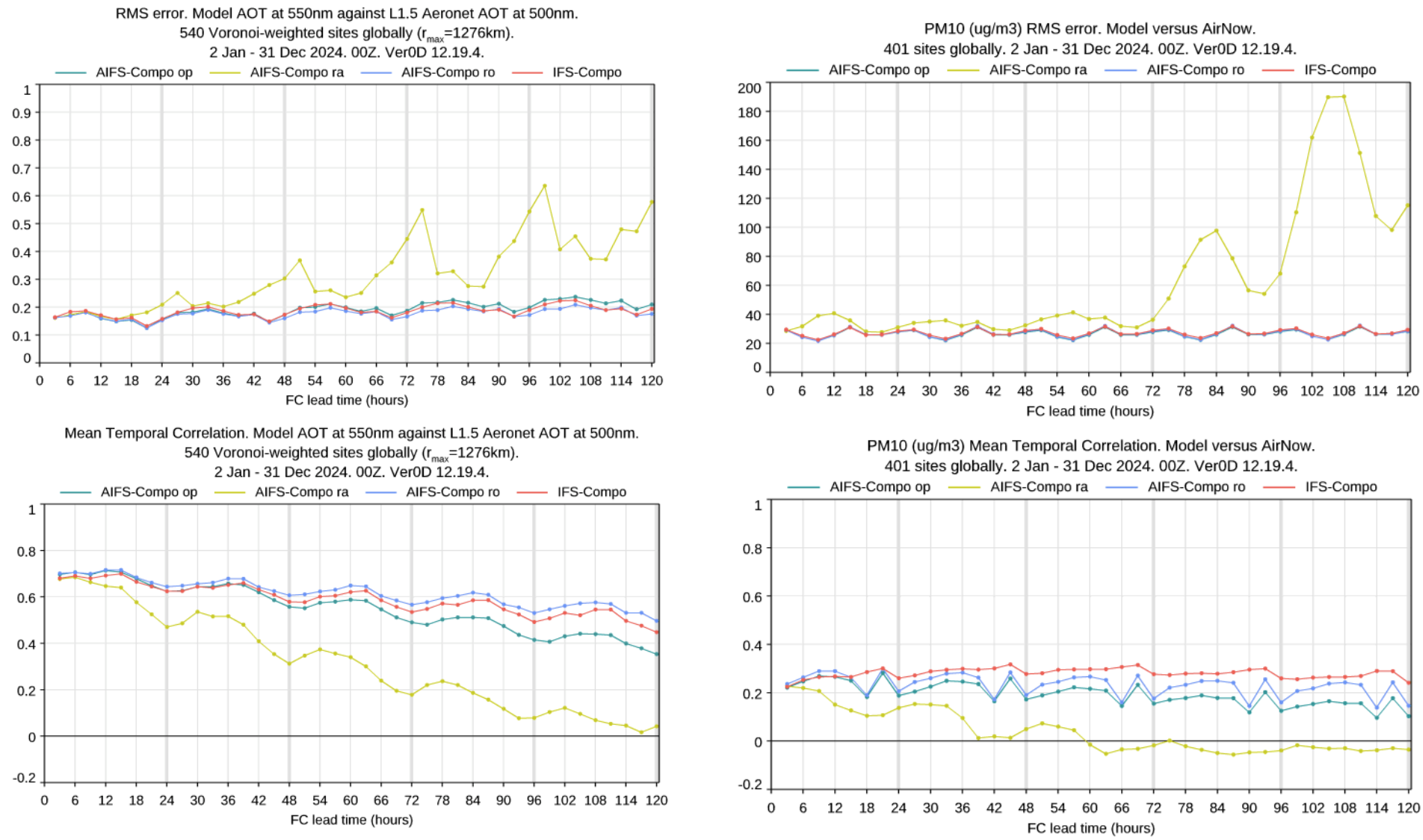}
\caption{RMSE (top) and temporal correlation (bottom) for AOD (left) and PM10 (right) compared against observations. It shows three different versions of AIFS-COMPO, each after a different training stage reanalysis pretraining (ra), opertional finetuning (op) and rollout finetuning (ro).}
\label{obs-stages}
\end{center}
\end{figure*}

To assess the contribution of the different training stages, we compare model performance after each step: AIFS-COMPO ra (reanalysis) after pretraining, AIFS-COMPO op (operational) after finetuning on operational data, and the final AIFS-COMPO ro (rollout) after rollout finetuning. Figure \ref{obs-stages} illustrates the substantial impact of these stages, in particular the pronounced reduction in RMSE following finetuning on operational data. The version trained solely on EAC4 reanalysis exhibits noticeably higher RMSE, with errors increasing from lead times of 18 hours for AOD and 6 hours for PM$_{10}$. For both variables, the RMSE grows rapidly with forecast lead time. A similar behavior is observed for the temporal correlation, which decreases sharply, reaching near-zero values by day 5 for AOD and already by day 2 for PM$_{10}$. 

Finetuning on operational data significantly improves performance compared to the reanalysis-only model, resulting in a much more stable RMSE across lead times. However, it still underperforms relative to IFS-COMPO. The final model, which additionally incorporates rollout finetuning, achieves the best overall performance across all metrics and variables. These results clearly demonstrate the importance of each training stage, with successive improvements at every step leading to substantial gains in forecast skill.

\subsection*{Evaluation against analysis}

We further evaluate AIFS-COMPO against global CAMS analysis data, using RMSE computed on the reduced Gaussian grid with equal weighting for each grid point. Due to the quasi-uniform area represented by grid points on this grid, this provides a consistent global assessment without explicit area weighting. The evaluation is performed over the full year 2024 in order to capture seasonal variability. Forecasts are initialized once per day at 00\,UTC and evaluated at 6-hourly intervals, consistent with the temporal availability of the analysis data. 
Overall, the results show strong performance of AIFS-COMPO for aerosol variables and many pressure levels, while performance is more mixed for particulate matter and weaker for total column quantities. This complementary evaluation provides insight into variables and regions not covered by observational datasets.

\subsubsection*{Aerosol Optical Depth}

\begin{figure*}[htb]
\begin{center}
\includegraphics[width=\columnwidth]{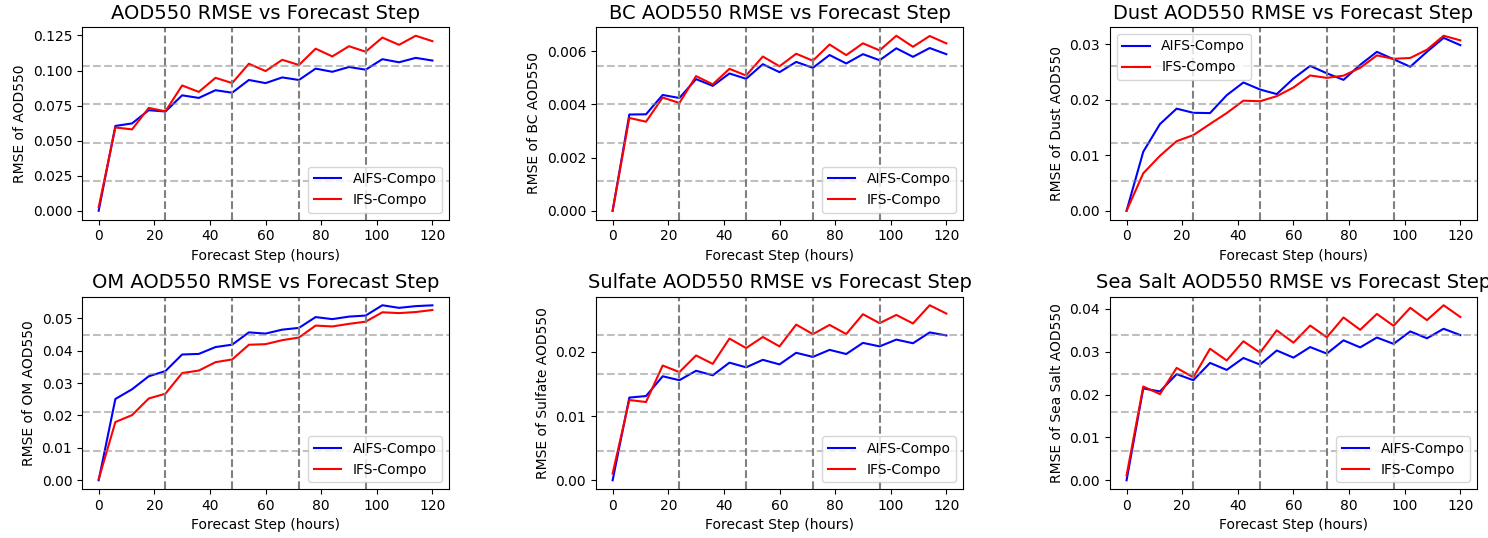}
\caption{RMSE of IFS-COMPO and AIFS-COMPO forecasts evaluated against CAMS analysis for total AOD at 550\,nm and its components: black carbon (BC), dust, organic matter (OM), sulphate, and sea salt.}
\label{aod-an}
\end{center}
\end{figure*}

Figure~\ref{aod-an} shows the RMSE of total AOD at 550\,nm and its components. AIFS-COMPO clearly outperforms IFS-COMPO for total AOD after the first forecast day. Similar improvements are observed for sulphate and sea salt AOD, while black carbon AOD shows a modest but consistent improvement from day 1 onward. For dust AOD, AIFS-COMPO does not match the performance of IFS-COMPO until approximately day 3. Organic matter (OM) AOD remains more challenging, with AIFS-COMPO showing higher RMSE throughout the forecast, although the gap decreases at longer lead times. Overall, AIFS-COMPO provides clear benefits for AOD, particularly at longer lead times. For all AOD variables, both models exhibit a pronounced oscillatory pattern in RMSE with a 6-hour periodicity.

\subsubsection*{Particulate Matter}

\begin{figure*}[htb]
\begin{center}
\includegraphics[width=\columnwidth]{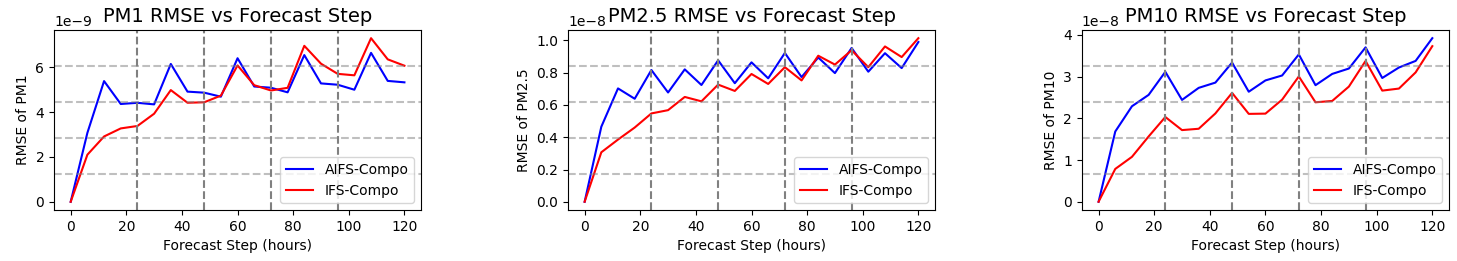}
\caption{RMSE of IFS-COMPO and AIFS-COMPO forecasts evaluated against CAMS analysis for PM$_1$, PM$_{2.5}$, and PM$_{10}$.}
\label{pm-an}
\end{center}
\end{figure*}

The evaluation of particulate matter (PM$_1$, PM$_{2.5}$, and PM$_{10}$) is shown in Figure~\ref{pm-an}. AIFS-COMPO exhibits higher RMSE at short lead times for all three variables. However, for PM$_1$ and PM$_{2.5}$, AIFS-COMPO surpasses IFS-COMPO after approximately 2–3 days, indicating improved performance at longer lead times. For PM$_{10}$, the RMSE gap between the two models decreases over time, but AIFS-COMPO remains slightly less accurate throughout the forecast horizon.

\subsubsection*{Total column variables}

\begin{figure*}[htb]
\begin{center}
\includegraphics[width=\columnwidth]{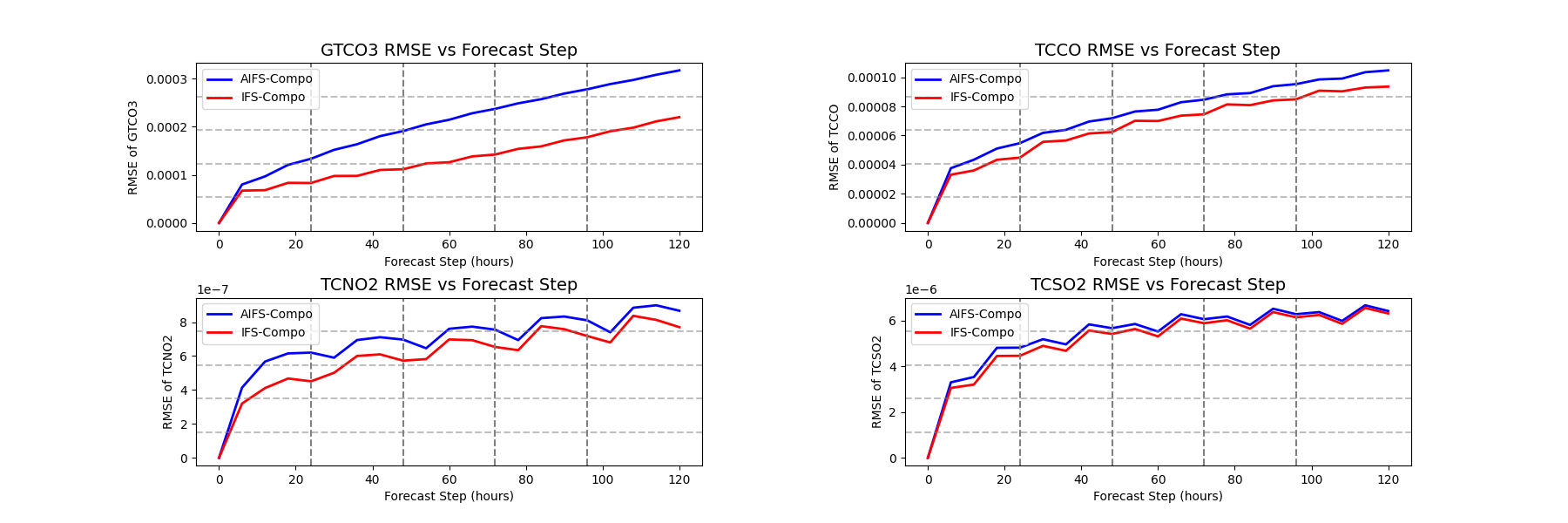}
\caption{RMSE of IFS-COMPO and AIFS-COMPO forecasts evaluated against CAMS analysis for total column ozone (O$_3$), CO, NO$_2$, and SO$_2$.}
\label{tc-an}
\end{center}
\end{figure*}

Figure~\ref{tc-an} presents results for total column ozone (O$_3$), carbon monoxide (CO), nitrogen dioxide (NO$_2$), and sulphur dioxide (SO$_2$). Total column variables represent a relative weakness of AIFS-COMPO compared to IFS-COMPO. In particular, total column ozone shows a substantial performance gap, with significantly higher RMSE for AIFS-COMPO. This degradation is primarily associated with the Northern Hemisphere, consistent with the good performance observed for Antarctic ozone hole predictions. For NO$_2$ and CO, AIFS-COMPO exhibits consistently higher RMSE across all lead times. For SO$_2$, performance is closer, with AIFS-COMPO approaching IFS-COMPO skill towards the end of the forecast horizon.

\subsubsection*{Pressure level variables}

\begin{figure*}[htb]
\begin{center}
\includegraphics[width=10cm]{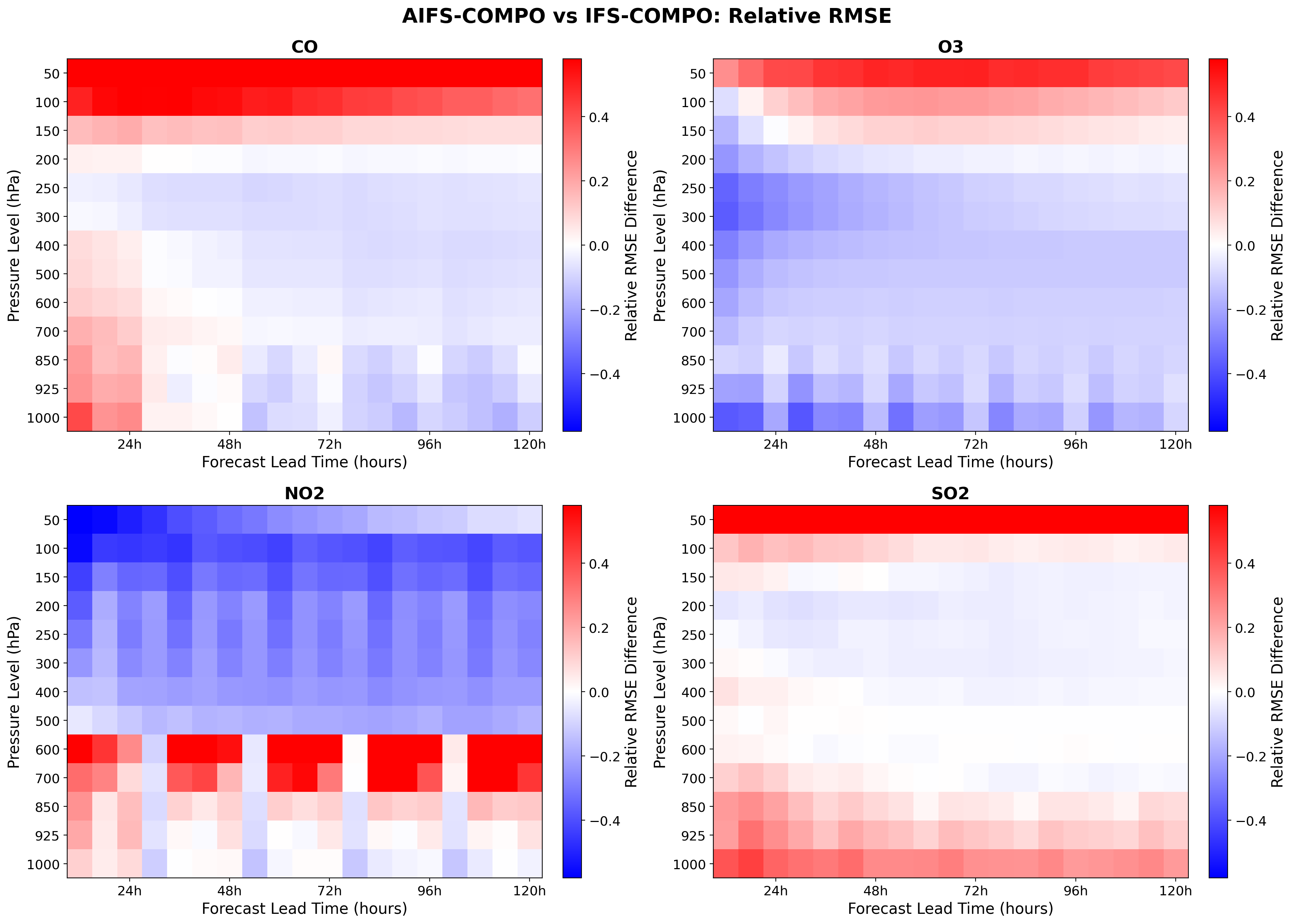}
\caption{Relative RMSE of IFS-COMPO and AIFS-COMPO forecasts evaluated against CAMS analysis for pressure-level predictions of CO, ozone, NO$_2$, and SO$_2$.}
\label{pl-an}
\end{center}
\end{figure*}

Figure~\ref{pl-an} summarizes the relative RMSE for pressure-level predictions of CO, ozone, NO$_2$, and SO$_2$. For ozone, AIFS-COMPO shows clear improvements over IFS-COMPO at mid- and lower tropospheric levels (1000–200\,hPa) across most lead times, particularly at short lead times. However, performance degrades in the stratosphere, with substantially higher RMSE at 50\,hPa. For CO AIFS-COMPO improves on IFS-COMPO for leadtimes after day 2 and pressure levls below 200hPa, but is far behind IFS-COMPO for the stratosphere. For NO$_2$, AIFS-COMPO performs better than IFS-COMPO at pressure levels above 500\,hPa for all lead times. Closer to the surface, a pronounced diurnal cycle is evident, with several time steps per day showing increased RMSE for AIFS-COMPO relative to IFS-COMPO. For SO$_2$, both models show similar performance in the mid-troposphere (700–150\,hPa). However, AIFS-COMPO exhibits larger errors near the surface and again at 50\,hPa.

\end{document}